\documentclass[%
 aps,
 jmp,%
 amsmath,amssymb,
 reprint,%
superscriptaddress,
]{revtex4-1}

 \usepackage{tikz,pgfplots}
\usepackage{dcolumn}
\usepackage{bm}
\usepackage{tabu}
\usepackage{calc}
\usepackage{xspace}

\newcommand{\beginsupplement}{%
        \setcounter{table}{0}
        \renewcommand{\thetable}{S\arabic{table}}%
        \setcounter{figure}{0}
        \renewcommand{\thefigure}{S\arabic{figure}}%
        \setcounter{equation}{0}
\setcounter{figure}{0}
\setcounter{table}{0}
\setcounter{page}{1}
     }

\makeatletter
\newcommand*{\balancecolsandclearpage}{%
  \close@column@grid
  \cleardoublepage
  \twocolumngrid
}
\makeatother

\graphicspath{{/}{figures/}}

\usepackage{animate}

\definecolor{fgred}{rgb}{0.6 ,0 ,0}
\definecolor{fgblack}{rgb}{0 ,0 ,0}

\newcommand{\ket}[1]{\left|#1\right\rangle}

\newcommand{\bra}[1]{\left\langle#1\right|}
\newcommand{\matrixel}[3]{\left< #1 \vphantom{#2#3} \right|
 #2 \left| #3 \vphantom{#1#2} \right>} 

\newcommand{\G}[1]{\(\Gamma_{#1}\)}

\newcommand{\azerox}{A$^0$X\xspace}
\newcommand{\azero}{A\(^0\)\xspace}
\newcommand{\bk}{\mathbf{k}\xspace}
\newcommand{\ephat}{\hat \epsilon\xspace}
\newcommand{\unit}[1]{\hat{\mathbf{#1}}}

\begin{document}
\title[]{Radiative properties of multi-carrier bound excitons in GaAs}

\author{Todd~Karin}
\author{Russell~Barbour}
\affiliation{Department of Physics, University of Washington, Seattle, Washington 98195, USA}
\author{Charles Santori}
\affiliation{Hewlett-Packard Laboratories, 1501 Page Mill Road, Palo Alto, California 94304, USA}
\author{Yoshihisa Yamamoto}
\affiliation{Edward L. Ginzton Laboratory, Stanford University, Stanford, California 94305-4085, USA}
\author{Yoshiro Hirayama}
\affiliation{Deptartment of Physics, Tohoku University, 6-3 Aramakiaza Aoba, Aobaku, Sendai, 980-8578 Japan}
\author{Kai-Mei C. Fu}
\affiliation{Department of Physics, University of Washington, Seattle, Washington 98195, USA}
\affiliation{Department of Electrical Engineering, University of Washington, Seattle, Washington 98195, USA}

\date{February 28, 2013}

\begin{abstract}
Excitons in semiconductors can have multiple lifetimes due to spin dependent oscillator strengths and interference between different recombination pathways. In addition, strain and symmetry effects can further
modify lifetimes via the removal of degeneracies. 
We present a convenient formalism for predicting the optical properties of \({k=0}\) excitons with an arbitrary number of charge carriers in different symmetry environments.
Using this formalism, we predict three distinct lifetimes for the neutral acceptor bound exciton in GaAs, and confirm this prediction through polarization dependent and time-resolved photoluminescence experiments. We find the acceptor bound-exciton lifetimes to be \({T_o (1,3,\frac{3}{4}  ) }\) where \({T_o = (0.61 \pm 0.12 ) \text{ ns}}\). Furthermore, we provide an estimate of the intra-level and inter-level exciton spin-relaxation rates.
\end{abstract}

\maketitle

The radiative properties of excitons in semiconductors are of fundamental interest in current semiconductor physics as well as of technological interest due to their impact on optoelectronic device performance~\cite{Chuang}. While the optical selection rules for the recombination of a conduction electron and valence hole are well understood~\cite{opticalOrientation,Chuang}, the selection rules for excitonic complexes with more than two carriers are complicated due to the multiple spin and angular degrees of freedom. In high symmetry environments, exciton lifetimes can be modified by interference between different recombination pathways~\cite{Efros1996}. This effect in quantum dots, for example, is responsible for the dark exciton, a radiative bottleneck in applications requiring bright sources~\cite{Donega2006} or alternatively a possible long-lived storage state for quantum information applications~\cite{Poem2010}.  Exciton lifetimes are also modified by reduced symmetry environments, which can remove the possibility of interference by energetically separating the excitonic states.

In this work, we provide a convenient and general framework for describing the optical properties of arbitrary \({k=0}\) excitonic complexes. We use the second quantization formalism~\cite{Kira} for calculating dipole matrix elements of an excitonic complex with an arbitrary number of electrons and holes in a III-V direct band gap semiconductor. Using a generalized Weisskopf-Wigner theory, we show how special spontaneous emission eigenstates and multiple radiative lifetimes may emerge. We predict three radiative lifetimes of the neutral acceptor bound-exciton (\azerox) in bulk GaAs. We confirm the theory by performing polarization dependent and time-resolved photoluminescence experiments on the \azerox system.

Exciton lifetimes in III-V direct band gap semiconductors can be derived from the dipole operator \({\boldsymbol \mu=e\mathbf r}\)  for band-to-band recombination between a \({j= \frac{1}{2}}\) conduction-band electron and a \({j=\frac{3}{2}}\) valence-band hole~\cite{Chuang,[][ pages 23-24] opticalOrientation}. In the second quantization formalism, the dipole operator is 
\begin{equation}\label{eq:dipoleOp}
\begin{aligned}
\boldsymbol\mu = \mu_o  & \left[ \frac{ \unit{x}+i \unit{y} }{\sqrt{2}}  \left(  h_{\frac{3}{2}} e_{-\frac{1}{2}}+ \frac{1}{\sqrt{3}} h_{\frac{1}{2}}  e_{\frac{1}{2}} \right)   \right. \\
& - \frac{ \unit{x}-i \unit{y}  }{\sqrt{2} }\left(h_{-\frac{3}{2}} e_{\frac{1}{2}} +\frac{1}{\sqrt{3} }  h_{-\frac{1}{2}} e_{-\frac{1}{2}}\right) \\
 & \left. + \sqrt{\frac{2}{3}}\unit{z} \left( h_{-\frac{1}{2}} e_{\frac{1}{2}} + h_{\frac{1}{2}} e_{-\frac{1}{2}}\right) + \text{H.C} \right] ,
\end{aligned}
\end{equation}
where  \(e_m\) (\(h_m\)) is the annihilation operator for an electron (hole) in the angular momentum state \(m\), H.C. is the Hermitian conjugate, and \(\mu_o\) is a spin-independent constant~\cite{A0Xsupp2}.
We define the coordinate system \(\unit x,\, \unit y \text{ and } \unit z\) to be oriented along the [100], [010] and [001] crystallographic directions. The hole angular momentum state is labeled with the opposite sign of the corresponding unoccupied electron angular momentum state. This dipole operator can be conveniently used to calculate the dipole matrix element between exciton states with an arbitrary number of charge carriers.
For example, the dipole matrix element \(\mathbf p_{ij}\) corresponding to the recombination of a two-carrier exciton with electron spin \(-\frac{1}{2}\) and hole spin~\(+\frac{3}{2}\) is
\[ \mathbf p_{0,\, e^\dagger_{-\frac{1}{2}}h^\dagger_{\frac{3}{2}} }  =  \bra 0  \boldsymbol \mu e^\dagger_{-\frac{1}{2}} h^\dagger_{\frac{3}{2}} \ket 0 = \mu_o \frac{ \unit{x}+i \unit{y} }{\sqrt{2}}, \]
where \(\ket 0\) is the semiconductor vacuum state.

We describe the radiative lifetimes of excitons using a generalized Weisskopf-Wigner theory. The spontaneous emission rates from a set of degenerate excited states to a set of degenerate ground states are the eigenvalues of \(\alpha S\), where \( S = \mathbf{p}^\dagger{\cdot}\mathbf{p}\)~\cite{A0Xsupp2}. Here \(\mathbf p\) is a matrix of the vector dipole matrix elements \(\mathbf p_{ij} = \bra{g,i} \boldsymbol \mu \ket{e,j} \) between ground state \(i\) and excited state \(j\), \(\alpha = (1/4\pi \epsilon)(4 \omega^3 n^3/3\hbar c^3  )\), \(\omega\) the frequency of the transition, \(\epsilon\) the permittivity of the material, \(n\) the index of refraction, and \(c\) the speed of light. The time dependence of the excited state probability amplitudes \(\mathbf b\) satisfy
\begin{equation}\label{eq:timedep}
 \frac{d}{dt} \mathbf b = - \frac{\alpha}{2} S\, \mathbf b ,
 \end{equation}
 corresponding to exponential decay.
Physically, Eq.~\ref{eq:timedep} implies that radiative lifetimes are modified by constructive or destructive interference between different recombination pathways. In addition, it highlights how exciton states organize into spontaneous emission eigenstates (eigenvectors of \(S\)) with decay rates given by the eigenvalues of \(\alpha S\).

Before applying this formalism to the three-carrier acceptor bound-exciton system, as an example we treat the simpler two-carrier light-hole exciton. Light-hole excitons, consisting of an ${m_j = \pm\frac12}$ valence hole and a conduction electron, split from heavy-hole excitons (${m_j = \pm\frac32}$) in reduced symmetry environments such as quantum dots, quantum wells and strained GaAs. The formalism yields the expected four spontaneous emission (SE) recombination rates~\cite{Efros1996}:
\begin{equation*}
  \arraycolsep=6pt\def\arraystretch{1.6}
 \begin{array}{cc}
 \text{SE Rate (normalized)} & \text{SE eigenstate} \\
 \hline
 \frac{1}{3} &  e^\dagger_{\frac{1}{2}} h^\dagger_{\frac{1}{2}} \ket 0 \\
  \frac{1}{3} &  e^\dagger_{-\frac{1}{2}} h^\dagger_{-\frac{1}{2}}   \ket 0\\
\frac{4}{3} &  \frac{1}{\sqrt 2} \left( e^\dagger_{\frac{1}{2}} h^\dagger_{-\frac{1}{2}} +e^\dagger_{-\frac{1}{2}} h^\dagger_{\frac{1}{2}} \right)  \ket 0\\
0 & \enspace  \frac{1}{\sqrt 2} \left( e^\dagger_{\frac{1}{2}} h^\dagger_{-\frac{1}{2}} -e^\dagger_{-\frac{1}{2}} h^\dagger_{\frac{1}{2}} \right)  \ket 0.
\end{array}
 \end{equation*}
The aligned spin \({e^\dagger_{\frac{1}{2}} h^\dagger_{\frac{1}{2}}}\) and  \({e^\dagger_{-\frac{1}{2}} h^\dagger_{-\frac{1}{2}}}\) excitons decay independently because of the orthogonal polarizations (\(\sigma^+\) and \(\sigma^-\)) of the transitions. On the other hand, constructive and destructive interference of the \(\unit z\) recombination pathway leads to a bright and dark exciton.
While it is experimentally challenging to observe the brightest light-hole exciton due its \(\unit z\) polarization, this exciton has recently been observed using magnetic-field measurements in strain-engineered quantum dots~\cite{Huo2014}.
We note that identification could alternatively be made through lifetime measurements at zero magnetic field.

\begin{figure}[hbt]
\includegraphics[width=3.4in]{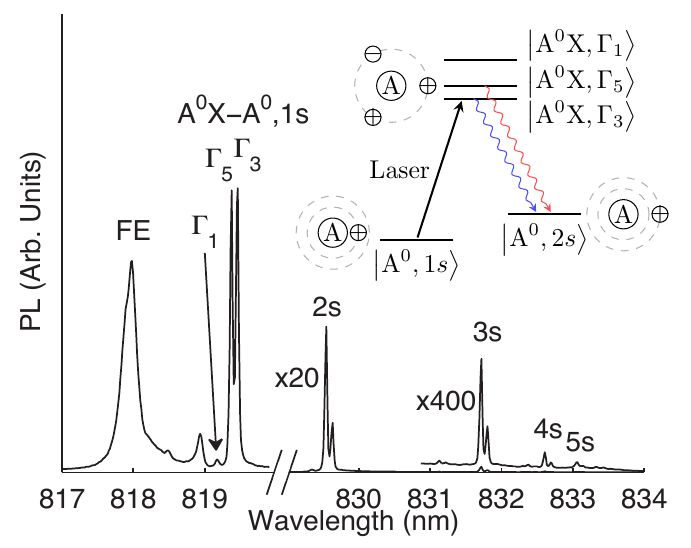} 
\caption{PL spectrum of \(\ket{\text{\azerox }}\to \ket{ \text{A}^0,n\text{s}}\) and free exciton (FE) transitions using above band and resonant excitation. The \azerox \G{3}-\G{5} splitting is clearly resolved. The spectrum left of the cut was taken using above band excitation at 815 nm. The spectrum right of the cut used resonant excitation of \(\ket{ \text{A}^0,1\text{s}} \to \ket{\text{\azerox },\Gamma_5}\). 
Because of spin relaxation between \azerox states, the different excitation conditions result in different \G{3} to \G{5} intensity ratios left and right of the cut. The inset shows an energy level diagram and cartoon of the \azero-\azerox system.  $T=2.3\,$K.}
\label{fig:hydrogenic}
\end{figure}

We now turn to the neutral acceptor-bound exciton~(\azerox), consisting of two \({j=\frac{3}{2}}\) holes and one \({j=\frac{1}{2}}\) electron bound to a substitutional acceptor impurity~\cite{Bogardus1968,Haynes1960}. By recombination of the electron with one of the holes, \azerox decays radiatively to a neutral acceptor (\azero, a hole bound to an acceptor). Effective mass theory can be used to show that \azero has hydrogenic levels \(1s\), \(2s\), etc. In high-purity p-type GaAs, \azerox to \azero 1s, 2s, etc. photoluminescence~(PL) is readily observed and provides a useful probe for resonant excitation, as shown in Fig.~\ref{fig:hydrogenic}. Remarkably, the ensemble transition linewidths of this solid-state ensemble system are less than 40 \(\mu\)eV (spectral linewidths in Fig.~\ref{fig:hydrogenic} are limited by the instrument resolution)~\cite{A0Xsupp2}.

\begin{figure}[hbt]
\includegraphics{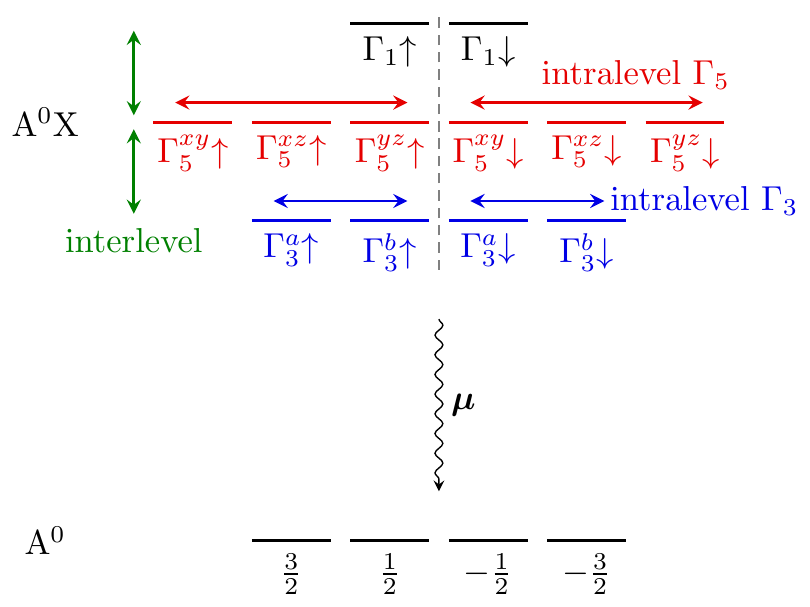}
\caption{Full energy level diagram for \azerox and \azero system. The spin of the electron is denoted by \(\uparrow\),\(\downarrow\). Intra-level and inter-level relaxation cause a decay of the polarization visibility (Fig.~\ref{fig:fits}b). Electron spin flips are not allowed, as depicted schematically by the dashed line. The crystal fields that split \G{3} and \G{5} lead to \(\unit x\), \(\unit y\), \(\unit z\) oriented along the crystallographic axes.
}
\label{fig:levelDiagram}
\end{figure}

Though the origin of the \azerox fine structure was once a controversy, strain experiments support hole-hole and crystal field coupling as the dominant mechanisms for splitting the 12 fold degenerate \azerox~\cite{Mathieu1984,Karasyuk1998}. In this scheme, the two holes lie in antisymmetric spin states with total spin 0 and 2~\cite{Mathieu1984}. Hole-hole coupling splits the \({j=0}\) states from \({j=2}\) states. In zinc-blende semiconductors which possess crystal fields with \(T_d\) symmetry, the \({j=2}\) states further split into two manifolds: \G{5} with multiplicity 3 and \G{3} with multiplicity 2. The full specification of \azerox also includes the spin of the electron, denoted as \(\uparrow\) or \(\downarrow\) (Fig.~\ref{fig:levelDiagram}).

Our theory and experiments show that \azerox has multiple radiative lifetimes. To find the \azerox radiative lifetimes, we first compute the dipole matrix elements \({\mathbf p_{ij} = \bra{\text{A}^0,i} \boldsymbol \mu \ket{\text{A}^0\text{X},j}}\) for the different \azero and \azerox states~\cite{A0Xsupp2}. The eigenvalues of \({\mathbf p^\dagger{\cdot}\mathbf p}\) are the radiative recombination rates of \azerox. An energy splitting between excited states causes a fast oscillation in the Weisskopf-Wigner theory which destroys the coupling between non-degenerate states~\cite{A0Xsupp2}. As such, only degenerate excited states are included in the dipole matrix when calculating the eigenvalues of \(\mathbf{p}^\dagger{\cdot} \mathbf{p}\).

\begin{figure*}[hbt]
\includegraphics{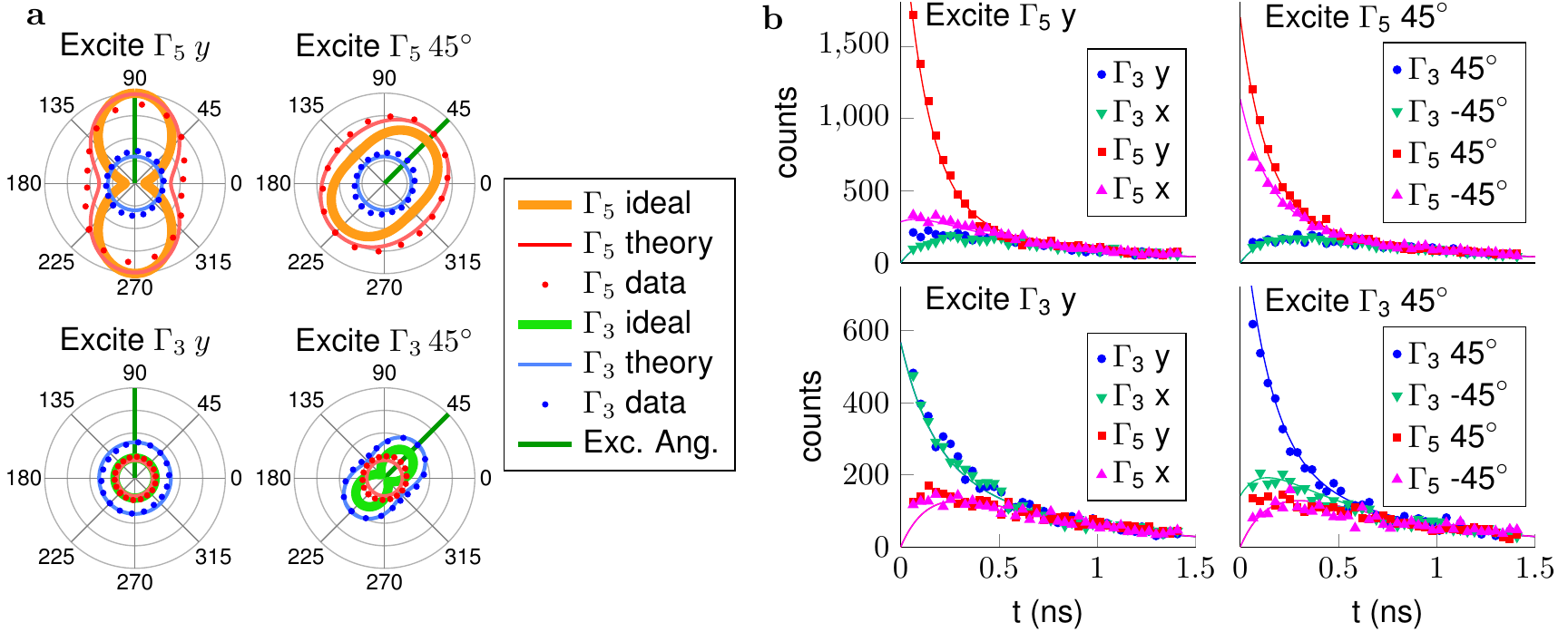}
\caption{
\textbf{a.} 
Dependence of light emitted from \G{3} and \G{5} on incident and collected polarization for \azerox \(\to\) \azero relaxation in resonant CW excitation. The dark green line shows the angle of excitation polarization, with horizontal \(\unit x\) corresponding to [100]. The polar plots show the PL emitted at the corresponding collection polarization angle.  The ideal theory curves depict the polarization dependence in the absence of excited state relaxation. The theory curves give the angular PL dependence expected from the density matrix model fit to the time-resolved data (Fig.~\ref{fig:fits}b). The data is PL collected from the \(\ket{\text{\azerox },\Gamma_n}\to \ket{ \text{A}^0,\text{2s}}\) transition while exciting the \(\Gamma_3\) or \(\Gamma_5\) line at \(\unit x\) or \(\frac{1}{\sqrt{2}} (\unit x - \unit y) = -45^\circ\) polarizations (Fig.~\ref{fig:hydrogenic}). All the data is normalized using the same constant. The normalization between the ideal curves and the theory curves is arbitrary. $T=2.3$ K.
\textbf{b.} The \G{3} and \G{5} PL as a function of time for polarizations \(\text{x}\propto[100]\), \(\text{y} \propto [010]\), \(45^\circ \propto [110]\) and \(\text{-45}^\circ \propto [1\bar{1}0]\) after an excitation pulse at \(t=0\). The lines are a simultaneous fit of the density matrix model to the 16 observed time dependent PL curves. Best fit parameters are given in the text.}
\label{fig:fits}
\end{figure*}

The \azerox spontaneous emission rates in spherical symmetry and no hole-hole spin coupling are proportional to 
 \[\overbrace{ \left(1,1, 2,2,2,2, \tfrac{1}{3},\tfrac{1}{3},\tfrac{1}{3},\tfrac{1}{3},\tfrac{1}{3},\tfrac{1}{3}  \right) }^{\frac{3}{2} \otimes \frac{3}{2} \otimes \frac{1}{2} }. \]
When hole-hole spin coupling is introduced, \({j=0}\) states split from \({j=2}\) states, but the spontaneous emission rates remain unchanged: \[ \overbrace{ (1,1) }^{j=0}, \overbrace{ \left( 2,2,2,2, \tfrac{1}{3},\tfrac{1}{3},\tfrac{1}{3},\tfrac{1}{3} ,\tfrac{1}{3} ,\tfrac{1}{3} \right)}^{j=2}.\]
With the inclusion of the zinc-blende crystal fields (which cause a \(\Gamma_5\)-\(\Gamma_3\) splitting), we find that the spontaneous emission rates become proportional to
\[
\overbrace{ (1,1)}^{\Gamma_1},
\overbrace{\left( 1,1,1,1 \right)}^{\Gamma_3},
\overbrace{\left( \tfrac{4}{3},\tfrac{4}{3},\tfrac{4}{3},\tfrac{4}{3},\tfrac{1}{3},\tfrac{1}{3} \right)}^{\Gamma_5}.
\]
Whereas previous studies of \azerox report only one lifetime~\cite{Hwang1973, Finkman1986} ($T_o = (1.6\pm0.6)$~ns), a full study including spin and symmetry shows that \azerox has multiple lifetimes differing by up to a factor of 4.
We can experimentally test this theory by studying the polarization dependence of photoluminescence (PL). If the system starts in an incoherent mixture of the four ground \azero states, excitation light of polarization \(\ephat_\text{i}\) resonant with \(\Gamma_n\) will create an excited state density matrix in the \(\Gamma_n\) subspace proportional to
\begin{equation}\label{eq:rhoExcite}
\rho_e^{(\Gamma_n)} = M_{\epsilon_\text{i}}^{(\Gamma_n)\dagger}  M_{\epsilon_\text{i}}^{(\Gamma_n)},
\end{equation}
where \(M_{\epsilon_\text{i}}^{(\Gamma_n)} = \mathbf{p}^{(\Gamma_n)} \cdot \ephat_\text{i}\), \(  \ephat_\text{i}\) is the incident polarization and \(\mathbf{p}^{(\Gamma_n)}\) are the dipole matrix elements corresponding to \(\Gamma_n\)~\cite{A0Xsupp2}. Eq.~\ref{eq:rhoExcite} is valid in the limit of low excited state population.
The PL emission from the states in \(\Gamma_n\) with polarization \(\epsilon_\text{f}\) is proportional to
\begin{equation}\label{eq:PLint}
\text{PL}^{(\Gamma_n)} = \text{tr} \left( M_{\epsilon_\text{f}}^{(\Gamma_n)} \rho_e^{(\Gamma_n)} M_{\epsilon_\text{f}}^{(\Gamma_n)\dagger}             \right).
\end{equation}
Eq.~\ref{eq:PLint} can be used to compute the arbitrary polarization dependence of \azerox-\azero transitions. In the case of exciting \G{n} with linear polarization at an angle \(\phi_\text{i}\) in the \(x\)-\(y\) plane and collecting linearly polarized light at \(\phi_\text{f}\) (\(\phi_n= 0\) corresponds to polarization along [100]), the angle dependent PL intensity is given by
\begin{equation*}
\begin{aligned}
\text{PL}^{(\Gamma_5)} &=  \frac{I_o}{18} [5 +  4 \cos(2 \phi_\text{i}) \cos(2 \phi_\text{f})  + \sin(2 \phi_\text{i})\sin(2\phi_\text{f}) ] \\
\text{PL}^{(\Gamma_3)} &= \frac{I_o}{36} \left[ 4 + 3 \sin ( 2 \phi_\text{i}) \sin (2 \phi_\text{f}) \right] \label{eq:G3xy} \\
\text{PL}^{(\Gamma_1)} &= \frac{I_o}{18} ,
\end{aligned}
\end{equation*}
where \(I_o\) is a constant.
These functions are plotted in Fig.~\ref{fig:fits}a for \G{3} and \G{5}. Here we note that this simple angular dependence of excitonic PL can be used to verify the relative inter-carrier and crystal field coupling for \azerox, once a subject of debate, without the need for applied strain or magnetic fields~\cite{Mathieu1984,Karasyuk1998,A0Xsupp2}. In the case where crystal fields have an observable effect, the excitonic PL can also be used to determine crystal orientation: e.g. \G{5} emission will be strongest when exciting along [100] and collecting [100].

We measure the polarization dependence of the ${\text{A}^0-\text{A}^0\text{X}}$ transition using resonant continuous-wave (CW) excitation.
Experiments were performed on a p-type GaAs crystal (6~\(\mu\)m GaAs grown by molecular beam epitaxy on a GaAs substrate, \({N_a = 1.2\cdot  10^{14}~\text{cm}^{-3}}\)). The sample was mounted without strain in pumped liquid He (1.9~K) and excited with a Ti:Sapphire laser. 
Fig.~\ref{fig:fits}a shows the polarization dependence of ${|\text{A}^0\text{X}\rangle \rightarrow |\text{A}^0, 2s\rangle}$ emission under resonant excitation of ${|\text{A}^0, 1s\rangle \rightarrow |\text{A}^0\text{X} \rangle}$ with $\unit y$ and $45^\circ$ polarized excitation.  The polarization visibility $ C =I_\text{max}/I_\text{min}$ observed is somewhat less than would be expected from the ideal theory. The difference can be explained by relaxation between \azerox spin states.

We investigate the effect of inter-level relaxation on the diminished polarization visibility with time- and polarization-resolved measurements. Either the \G{5} or \G{3} transitions were excited resonantly with 2~ps Ti:Sapphire pulses, spectrally filtered to obtain 16~ps pulses with 0.03~nm bandwidth. Photoluminescence to ${|\text{A}^0, 2s\rangle}$ was collected and imaged using a combined spectrometer/streak camera setup with a timing resolution of 27~ps. Four excitation conditions were studied, resonant excitation of \G{5} or \G{3} with $\hat{\mathbf{y}}$ or 45$^\circ$ linearly polarized light. PL polarized parallel and perpendicular to the excitation polarization was collected. The complete time-resolved data set is shown in Fig.~\ref{fig:fits}b.

We observe a strong initial polarization visibility at \({t=0}\) that later decays because of inter-level relaxation. The initial polarization visibility of \({C = 7.2 \pm 1}\) for \({\Gamma_5-y}\) excitation is close to the ideal value 9 for no excited state relaxation. (The uncertainty here is due to the uncertainty in the \({t=0}\) time.) The decay of polarization visibility indicates the existence of spin flip processes on the same timescale as the radiative lifetime.

We use the time-resolved data to obtain estimates of the inter- and intra-level relaxation rates in the exciton system (Fig.~\ref{fig:fits}b). The time resolved data were fit to a 12 state density matrix model including inter-state relaxation~\cite{A0Xsupp2}. In the model, an optical pulse of a given polarization coherently creates an excited state density matrix given by Eq.~\ref{eq:rhoExcite}. The subsequent time evolution of the excited state density matrix \(\rho\) satisfies
\begin{equation*}
 \frac{d\rho}{dt}= \frac{1}{i \hbar} [H_o,\rho] - \frac{1}{2} \left\{\rho, \alpha \mathbf p^\dagger{\cdot}\mathbf p \right\} + L(\rho) ,
 \end{equation*}
 where \(H_o\) is a diagonal matrix of the excited state energies, the second term describes radiative recombination and \(L(\rho)\) is the Linbladian operator describing phenomenological relaxation between excited states~\cite{A0Xsupp2}. From the solution \(\rho(t)\) we calculate the relative PL intensity emitted into different polarizations using Eq.~\ref{eq:PLint}.

This model gives a good fit to the observed time dependence of \azerox emission (Fig.~\ref{fig:fits}b). The 16 curves in Fig.~\ref{fig:fits}b are fit simultaneously using 6 fit parameters: overall spontaneous emission rate (1.48~ns$^{-1}$), inter-level relaxation (0.89~ns$^{-1}$), intralevel \G{3} relaxation (3.6~ns$^{-1}$), intralevel \G{5} relaxation (1.8~ns$^{-1}$), temperature (4.7 K) and overall intensity normalization (5100 counts). These relaxation rates are shown schematically in Fig~\ref{fig:levelDiagram}.
 The resulting  \azerox radiative lifetimes are \(T_o (1,3,\frac{3}{4}  )  \) where \(T_o\) lies in the range 0.49 to 0.74 ns. A detailed error analysis found the main source of uncertainty in the spontaneous emission rate to be due to an ambiguity in the choice of the background level~\cite{A0Xsupp2}.  Since hole spin flips are predicted to be much faster than electron spin flips~\cite{Fu2006,Kroutvar2004}, we do not include electron spin flip processes in the model (shown schematically in Fig.~\ref{fig:levelDiagram}). Temperature was included as a fit parameter because the effective temperatures for bound excitons can be larger than the bath temperature~\cite{Ruhle1978}.
 
Using the best fit model parameters from the time resolved experiment, we are now able to predict the polarization dependence of PL in resonant CW excitation in the presence of spin relaxation. These curves are shown in Fig.~\ref{fig:fits}a as ``theory," and agree well with the experimental data.

In conclusion, we presented a convenient and general formalism for calculating the optical properties of \({k=0}\) excitons in III-V semiconductors with an arbitrary number of carriers.
We used this formalism to derive a model of the optical properties of \azerox in strain-free bulk GaAs which predicts 3 distinct radiative lifetimes. The model was confirmed using polarization and time-resolved experiments.  The results are in contrast to previous reports for this system and highlight the importance of a unified treatment of all recombination pathways when deriving the radiative properties of multi-carrier excitons.   



This material is based upon work supported by the National Science
Foundation under Grant No. 1150647, DGE-0718124 and DGE-1256082. We would like to thank T. Saku for growing the material in NTT.
YH acknowledges support from SORST and ERATO programs by JST.

\bibliography{toddkarin.bib}

\balancecolsandclearpage

\beginsupplement
\pagebreak
\onecolumngrid 
\begin{center}
\textbf{\large Supplemental Materials:\\Radiative properties of multi-carrier bound excitons in GaAs} \\
\vspace{1cm}
\end{center}
\twocolumngrid

\section{Dipole Operator} \label{sec:dipoleOp}

The vector dipole operator \(\bm \mu = e \mathbf r\) for transitions between the conduction band and the heavy-hole or light-hole band in a zinc-blende direct band gap semiconductor (e.g. GaAs, InP, etc.) can be derived from the electron and hole basis functions. We derive the dipole operator in second quantization, which is convenient for calculating recombination rates for excitons with more than two charge carriers.

The valence band angular momentum states arise from coupling between p-like orbital states and the electron spin \(\frac{1}{2}\)~\cite{Yu}. These couple together to form total angular momentum \(\frac{3}{2}\) and \(\frac{1}{2}\):
\[ 1\otimes \frac{1}{2} = \frac{3}{2} \oplus \frac{1}{2} .\]
The \(\ket{\frac{3}{2}, \pm \frac{3}{2} }\) are known as heavy holes, \(\ket{\frac{3}{2}, \pm \frac{1}{2} }\) as light holes and \(\ket{\frac{1}{2}, \pm \frac{1}{2}} \) as split-off holes. Spin-orbit interaction splits the \({j=\frac{3}{2}}\) from the \({j=\frac{1}{2}}\) states and typically the \(j=\frac{1}{2}\) split-off holes can be ignored in experiments.

Using angular momentum addition rules, the heavy hole and light hole states are
\begin{equation*}
\begin{aligned}
\ket{\tfrac{3}{2}} &= \frac{1}{\sqrt 2} \ket{X-iY, \downarrow} \\
\ket{{\tfrac{1}{2}}} &= \frac{1}{\sqrt 6} \ket{X-iY, \uparrow} + \sqrt{\frac{2}{3}} \ket{Z, \downarrow} \\
\ket{{-\tfrac{1}{2}}} &= -\frac{1}{\sqrt 6} \ket{X+iY, \downarrow} +\sqrt{\frac{2}{3}} \ket{Z, \uparrow} \\
\ket{{-\tfrac{3}{2}}} &= -\frac{1}{\sqrt 2} \ket{X+iY, \uparrow}
\end{aligned}
\end{equation*}
where \(X,Y,Z\) are electron orbital wave functions transforming as \(x,y,z\) and \(\uparrow,\downarrow\) is the spin of the electron~\cite{Chuang}. The hole angular momentum state has the opposite sign of the corresponding electron angular momentum.
The conduction band states are \( \ket{S,\uparrow}\) and \(\ket{S,\downarrow}\) where \(S\) denotes a spherically symmetric periodic part of the Bloch wave function. 
In spherical symmetry and for a \({ k=0}\) exciton, the coordinate system can be taken to lie in an arbitrary direction. However for an exciton with non-zero momentum \(\mathbf k\), the coordinate system must be taken with the z axis in the \(\mathbf k\) direction, thus somewhat complicating further analysis~\cite{Chuang}. In what follows, we will restrict our discussion to \(k=0\) excitations.

These basis functions can be used to calculate matrix elements of the dipole operator \(\bm \mu\).
As an example, the dipole matrix element for recombination of a spin down electron with a \(+\frac{3}{2}\) heavy hole is
\begin{equation*}
\begin{aligned}
\matrixel{\tfrac{3}{2}}{\bm \mu}{S,\downarrow} &= \frac{e}{\sqrt{2}} \left(\bra{X} + i \bra{Y} \right) \mathbf r \ket{S} \langle \downarrow | \downarrow \rangle \\
										&= \frac{e}{\sqrt{2}} \left[ \bra{X} x\ket{S} \unit x+ i  \bra{Y} y \ket{S} \unit y \right] .\\
 \end{aligned}
 \end{equation*}
The ordering of the matrix element \(\bra f V \ket i\) reflects the transition occurring, in this case an electron moving from the conduction band to the valence band.
 In a bulk cubic crystal, by symmetry the matrix elements
  \[\bra{X} x\ket{S} = \bra{Y} y\ket{S} = \bra{Z} z\ket{S} \equiv \frac{1}{e} \mu_o \]
  are all identical. Further simplifying, we find this transition results in the production of right handed circularly polarized light:
\begin{equation*}
\matrixel{\tfrac{3}{2}}{\bm \mu}{S,\downarrow}   = \mu_o \frac{ \unit x+ i \unit y }{\sqrt 2}.
 \end{equation*}
 The same procedure can be used to find the other dipole matrix elements.

We can now introduce creation operators (\(e^\dagger_m\))~\cite{Kira,Knox} for the creation of an electron in the angular momentum state \(m\) and some particular but unspecified spatial state.
Because of the anti-symmetrization requirement, the creation and annihilation operators satisfy anti-commutation relations
\begin{equation}
\label{eq:commutations}
\begin{aligned}
\{ e_m , e_n^\dagger \} & = \delta_{mn} \\
\{ e_m^\dagger , e_n^\dagger \} & = 0  \\
\{ e_m , e_n\} & = 0.
\end{aligned}
\end{equation}
where the anti-commuator is defined as \(\{a,b\} = ab+ba\).
We will also introduce creation operators for holes using \(h^\dagger_{m, \bk} = e_{-m, -\bk}\); i.e. the linear and angular momentum of the hole has the opposite sign of the unfilled electron state~\cite{Kira}. Instead of labeling the band index, we restrict hole creation/annihilation operators to act in the valence band and electron operators in the conduction band.
For example, an exciton state can be written as 
\[ \ket{e_{m_1},h_{m_2}} = e^\dagger_{m_1} h^\dagger_{m_2} \ket 0 .\]
where \(\ket 0\) is the semiconductor vacuum state with a filled valence band and empty conduction band.

\begin{figure}[hbt]
\includegraphics{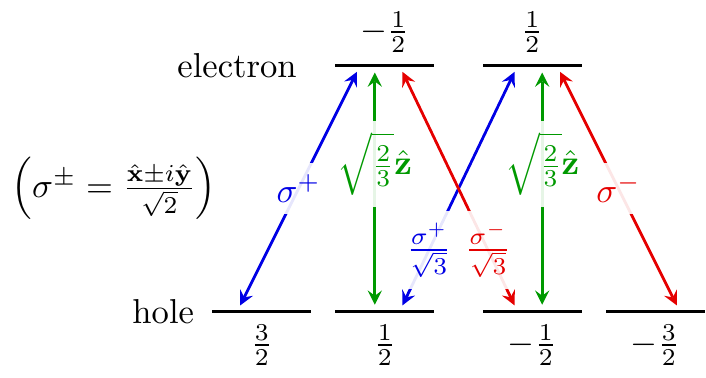}
\caption{Dipole matrix elements for exciton recombination~\cite{opticalOrientation}. The labeled arrows show the polarization of light emitted upon recombination of the corresponding electron-hole pair. }
\label{fig:exciton}
\end{figure}

The dipole operator can be written in second quantization as 
\begin{equation}\label{eq:dipole2q}
 \boldsymbol\mu = \sum_{mn} \boldsymbol\mu_{mn} h_m e_n + \boldsymbol\mu_{mn}^* e^\dagger_n h^\dagger_m 
 \end{equation} where we have restricted \(m\) to be in the valence band and \(n\) to be in the conduction band, and \(\boldsymbol\mu_{mn} = \bra m e \mathbf r \ket n \)~\cite{Kira}. The first term corresponds to exciton annihilation and the second to exciton creation.  Using the matrix elements calculated above, the dipole operator for a conduction band electron recombining with a heavy-hole or light-hole is

\begin{equation}\label{eq:dipoleOp}
\begin{aligned}
\boldsymbol\mu = \mu_o  & \left[ \frac{ \unit{x}+i \unit{y} }{\sqrt{2}}  \left(  h_{\frac{3}{2}} e_{-\frac{1}{2}}+ \frac{1}{\sqrt{3}} h_{\frac{1}{2}}  e_{\frac{1}{2}} \right)   \right. \\
& - \frac{ \unit{x}-i \unit{y}  }{\sqrt{2} }\left(h_{-\frac{3}{2}} e_{\frac{1}{2}} +\frac{1}{\sqrt{3} }  h_{-\frac{1}{2}} e_{-\frac{1}{2}}\right) \\
 & \left. + \sqrt{\frac{2}{3}}\unit{z} \left( h_{-\frac{1}{2}} e_{\frac{1}{2}} + h_{\frac{1}{2}} e_{-\frac{1}{2}}\right) + \text{H.C} \right] .
\end{aligned}
\end{equation}
Each term in the dipole operator (Eq.~\ref{eq:dipoleOp}) conserves angular momentum; i.e., the total electron and hole spin z projection is transferred to the photon during recombination. The dipole operator is shown schematically in Fig.~\ref{fig:exciton}.

\section{Basis states for \azerox}\label{app:groupTheory}

The \azerox consists of two holes and one electron. Hole-hole coupling dominates, while the crystal fields split the levels further~\cite{Mathieu1984}. From the two holes, there are four possible total spin states: \(\frac{3}{2} \otimes \frac{3}{2} = 0  \oplus 1 \oplus 2 \oplus 3\)~\cite{Sakurai}. The two holes in \azerox lie in a symmetric spatial state. On account of the Pauli principle, the spin state must therefore be antisymmetric with respect to interchange, resulting in only total spin 2 and 0 being allowed.

\begin{figure}[hbt]
\includegraphics[width=3.4in]{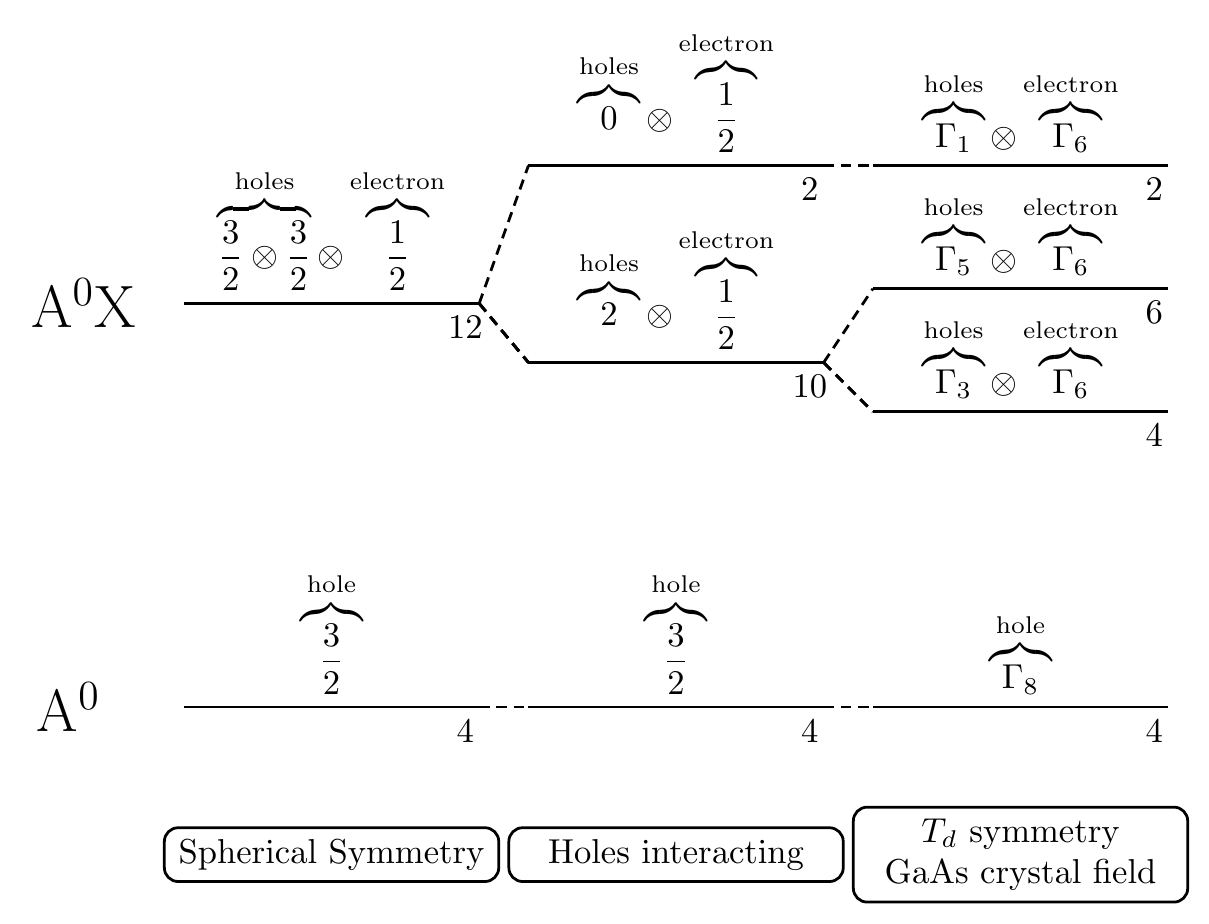}
\caption{Hole-hole coupling causes the \(j=2\) states to be split from the \(j=0\) states. Crystal fields breaking spherical symmetry and split \(\Gamma_3\) and \(\Gamma_5\). The degeneracy of the level is shown in the bottom right corner.}
\label{fig:splitting}
\end{figure}

We will use
\begin{equation}\label{eq:defcreate}
h^\dagger_{m_1} h^\dagger_{m_2} \ket 0 = \frac{1}{\sqrt 2} \left( \ket{m_1,m_2} - \ket{m_2,m_1} \right)
\end{equation}
as shorthand for the creation of an antisymmetric state of two holes~\cite{Kira}. Note that the ordering of the creation operators matters, consistent with the commutation relations in Eq.~\ref{eq:commutations}. Using this notation, we can write the total  angular momentum states \(\ket{j,m}\) for the coupling of the two holes as:

\begin{equation*}
\begin{aligned}
  \ket{2,2} &= h^\dagger_\frac{3}{2} h^\dagger_\frac{1}{2} \ket 0 \\
    \ket{2,1} &= h^\dagger_\frac{3}{2} h^\dagger_{-\frac{1}{2}} \ket 0 \\
    \ket{2,0} &= \frac{1}{\sqrt 2} \left( h^\dagger_\frac{3}{2} h^\dagger_{-\frac{3}{2}} + h^\dagger_\frac{1}{2} h^\dagger_{-\frac{1}{2} }\right) \ket 0 \\
    \ket{2,-1} &= h^\dagger_\frac{1}{2} h^\dagger_{-\frac{3}{2}} \ket 0 \\
    \ket{2,-2} &= h^\dagger_{-\frac{1}{2}} h^\dagger_{-\frac{3}{2}} \ket 0 \\
    \ket{0,0}  &= \frac{1}{\sqrt 2} \left( h^\dagger_\frac{3}{2} h^\dagger_{-\frac{3}{2}} - h^\dagger_\frac{1}{2} h^\dagger_{-\frac{1}{2} }\right) \ket 0 .
    \end{aligned}
 \end{equation*}
In zince-blende semiconductors, hole-hole coupling splits the \(j=2\) and \(j=0\) states (Fig.~\ref{fig:splitting}).

In the presence of the crystal field with \(T_d\) symmetry, the states split into three different irreducible representations~\cite{Mathieu1984,Dresselhaus,Francoeur2010} (Fig.~\ref{fig:splitting}):
\begin{equation*}
\begin{aligned}
 \ket{\Gamma_5^{xy}} &= \frac{1}{\sqrt{2}} \left( \ket{2,2} - \ket{2,-2} \right) \\
 \ket{\Gamma_5^{xz}} &= \frac{1}{\sqrt{2}} \left( \ket{2,1} +  \ket{2,-1} \right) \\
  \ket{\Gamma_5^{yz}} &= \frac{1}{\sqrt{2}} \left( \ket{2,1} -  \ket{2,-1} \right) \\
   \ket{\Gamma_3^{a}} &= \frac{1}{\sqrt{2}} \left( \ket{2,2} +\ket{2,-2} \right) \\
 \ket{\Gamma_3^{b}} &=  \ket{2,0} \\
  \ket{\Gamma_1} &= \ket{0,0}.
\end{aligned}
\end{equation*}
In order to derive these basis states, it is necessary to choose a coordinate system in which to write the symmetry operations of the crystal. Since we chose to use axes aligned along [100], [010] and [001], \(\unit x\), \(\unit y\) and \(\unit z\) correspond to the three crystallographic directions.

\section{Dipole Matrix Elements of \azerox}\label{sec:dipoleMatrixElements}

We use the dipole operator \eqref{eq:dipoleOp} to calculate the dipole matrix element between \azerox  and \azero. To illustrate the method, we will calculate the matrix element \( \bra{ \frac{3}{2} }\boldsymbol \mu \ket{\Gamma_5^{xy}\uparrow } \) as an example. First, we expand the matrix element and the dipole operator
\begin{equation*}
\begin{aligned}
\bra{ \tfrac{3}{2} }\boldsymbol \mu \ket{\Gamma_5^{xy}\uparrow } &= \mu_o \bra{0} h_\frac{3}{2}  \cdot \\
  & \left[ \frac{ \unit{x}+i \unit{y} }{\sqrt{2}}  \left(  h_{\frac{3}{2}} e_{-\frac{1}{2}}+ \frac{1}{\sqrt{3}} h_{\frac{1}{2}}  e_{\frac{1}{2}} \right)   \right. \\
& - \frac{ \unit{x}-i \unit{y}  }{\sqrt{2} }\left(h_{-\frac{3}{2}} e_{\frac{1}{2}} +\frac{1}{\sqrt{3} }  h_{-\frac{1}{2}} e_{-\frac{1}{2}}\right) \\
 & \left. + \sqrt{\frac{2}{3}}\unit{z} \left( h_{-\frac{1}{2}} e_{\frac{1}{2}} + h_{\frac{1}{2}} e_{-\frac{1}{2}}\right) \right] \cdot \\
& \frac{1}{\sqrt{2}}  e^\dagger_\frac{1}{2} \left( h^\dagger_\frac{3}{2}  h^\dagger_\frac{1}{2} -h^\dagger_{-\frac{1}{2}}  h^\dagger_{-\frac{3}{2}} \right) \ket{0}.
\end{aligned}
\end{equation*}
All terms with an electron annihilation operator \( e_{-\frac{1}{2}}\) go to zero because the electron in \(\ket{\Gamma_5^{xy}\uparrow }\) is spin up. Using the fact that \(  e_{\frac{1}{2}}e^\dagger_{\frac{1}{2}} \ket 0 = \ket 0\), the expression becomes
\begin{equation*}
\begin{aligned}
\bra{ \tfrac{3}{2} }\boldsymbol \mu \ket{\Gamma_5^{xy}\uparrow } = &\bra{0} h_\frac{3}{2}  \left[ \frac{ \unit{x}+i \unit{y} }{\sqrt{6}}h_{\frac{1}{2}}  -\frac{ \unit{x}-i \unit{y}  }{\sqrt{2}}h_{-\frac{3}{2}}  + \right. \\
 & \left.  \sqrt{\frac{2}{3}}\unit{z}  h_{-\frac{1}{2}} \right]   \frac{1}{\sqrt{2}}  \left( h^\dagger_\frac{3}{2}  h^\dagger_\frac{1}{2} -h^\dagger_{-\frac{1}{2}}  h^\dagger_{-\frac{3}{2}} \right) \ket{0}.
\end{aligned}
\end{equation*}
Using the fact that \(h_m \ket 0 = 0\), and the commutation relations in Eq.~\ref{eq:commutations}, the dipole matrix element is
\begin{equation*}
\bra{ \tfrac{3}{2} }\boldsymbol \mu \ket{\Gamma_5^{xy}\uparrow } = - \frac{ \unit{x}+i \unit{y} }{2\sqrt{3}}.
\end{equation*}
Repeating this calculation for each matrix element produces the dipole matrix elements for the \azerox-\azero system, given in Table~\ref{tab:dipoleMat}.

\begin{table*}[htbp]
\includegraphics{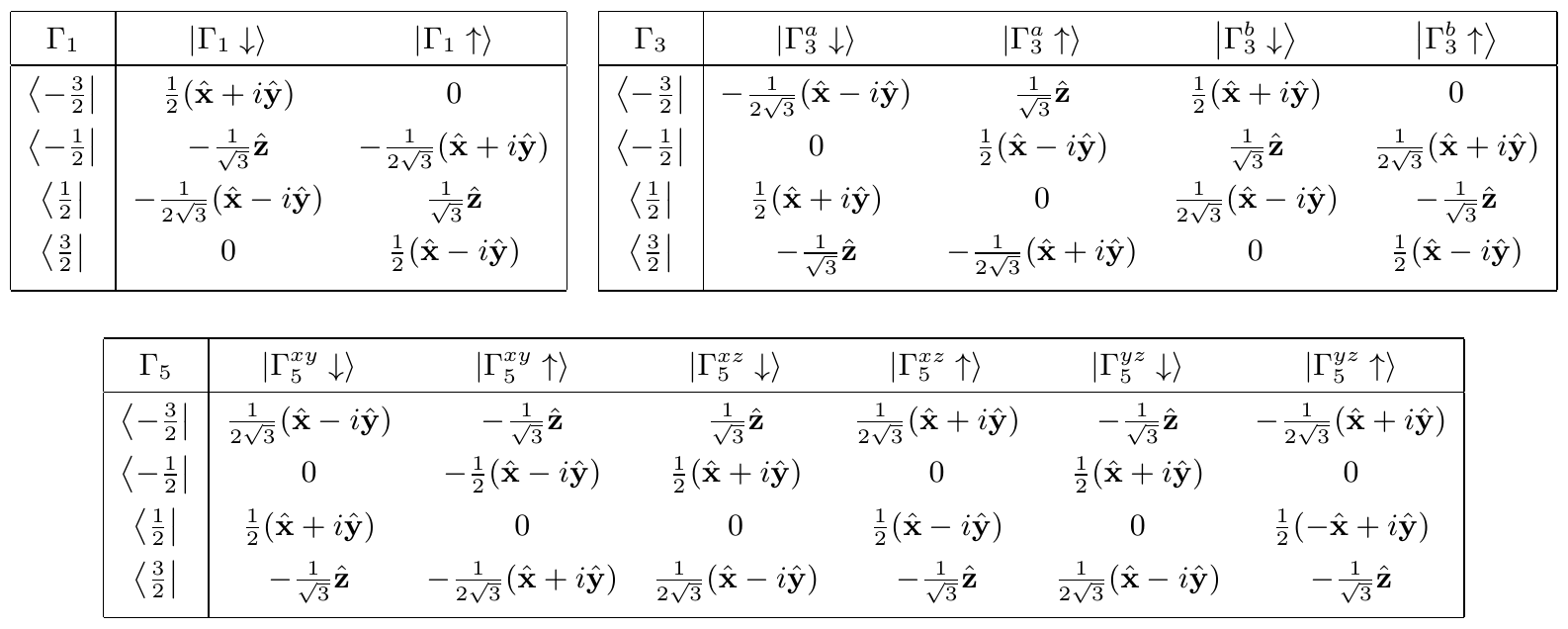}
\caption{Normalized dipole matrix \(\mathbf{p}_{ij} = \bra{ \text{A}^0, i} \boldsymbol \mu \ket{ \text{A}^0 \text{X},j }\) for the \azerox-\azero system. \(\unit{x}\), \(\unit{y}\) and \(\unit{z}\) are unit vectors oriented along the crystallographic axes.}\label{tab:dipoleMat}
\end{table*}

\section{Generalized Weisskopf-Wigner theory for spontaneous emission from multiple excited levels}\label{sec:WW}

The Wiesskopf-Wigner theory of spontaneous emission~\cite{Scully,Berman,Hindmarsh} can be generalized to calculate the spontaneous emission rate from a set of excited states to a set of ground states. The excited and ground states are not necessarily degenerate.

The wavefunction of the system is
\begin{equation}\label{eq:psi}
\ket{\psi(t)} = \sum_i \sum_{\bk} c_{i\bk}(t) \ket{g,i,\bk}  + \sum_j b_j(t) \ket{e,j}
\end{equation}
where \(  \ket{g,i,\bk}\) is the state with the atom in ground state~\(i\) and a photon in mode \(\bk\), polarization \(\epsilon_\sigma\) (\(\sigma = 1,2\)), and \(\ket{e,j} \) is the \(j\text{'th}\) excited atomic state and no photon. The sum on \(\bk\) contains an implicit sum over the two polarizations \(\sigma = 1,2\).

In the interaction picture and rotating wave approximation, the Hamiltonian governing the time evolution of the atom and field is
\begin{equation}\label{eq:V}
\mathcal V = \hbar \sum_{ij}\sum_{\mathbf k} \left[ g_\mathbf{k}^{ij}(\mathbf r_o)^* \ket{e,j} \bra{g,i} a_\mathbf{k} e^{i (\omega_{ij} - \nu_\mathbf{k} )t} + \text{H.C.} \right]
\end{equation}
where
\begin{equation*}
g_\mathbf{k}^{ij}(\mathbf r_o) = - \frac{\mathbf{p}_{ij} \cdot \ephat}{\hbar} \sqrt{ \frac{\hbar \nu_k}{2 \epsilon V} },
\end{equation*}
 \(\mathbf{p}_{ij}=\bra{g,i} \boldsymbol{\mu}\ket{e,j} \) are dipole matrix elements, \(a_\bk\) is the annhilation operator for a photon in mode \(\bk\), \(\omega_{ij} = (E_j - E_i)/\hbar \), \(\ephat\) is the polarization of the photon,  \(\omega\) is the transition frequency, \(\nu_\bk = c |\bk| /n\), \(\epsilon\) is the material permittivity, \(n\) is the material dielectric constant, \(c\) is the speed of light, and \(V\) is the quantization volume~\cite{Scully}. 
 
The time evolution is given by the Schr\"{o}dinger equation
\begin{equation*}
\frac{d}{dt} \ket{\psi(t)} =  - \frac{i}{\hbar} \mathcal{V} \ket{\psi(t)}.
\end{equation*}
This yields the coupled differential equations
\begin{equation}\label{eq:diffeq}
\begin{aligned}
\dot{b}_{j}(t) & = - i \sum_i \sum_\bk  g_\mathbf{k}^{ij}(\mathbf r_o) c_{i\bk}(t) e^{-i (\omega_{ij} - \nu_\bk)t} \\
\dot{c}_{i\bk}(t) & = - i \sum_j  g_\mathbf{k}^{ij}(\mathbf r_o)^* b_{j}(t) e^{i (\omega_{ij} - \nu_\bk)t}.
\end{aligned}
\end{equation}
By formally solving the second equation and plugging into the first, the time evolution of the excited state probability amplitude satisfies
\begin{equation}\label{eq:eqmotion}
\begin{aligned}
\dot{b}_{j'}(t) =   - \sum_{ij}& \sum_\bk   g_\mathbf{k}^{ij'}(\mathbf r_o) g_\mathbf{k}^{ij}(\mathbf r_o)^*   \int_o^t dt'   \\
&b_j(t') e^{-i (\omega - \nu_\bk)(t-t')+ i \Delta_{ij}t'- i \Delta_{ij'}t } ,
\end{aligned}
\end{equation}
where \(\Delta_{ij} = \omega_{ij}-\omega \) and \(\omega\) is some choice of natural frequency for the system. Assuming the modes are closely spaced in frequency, the sum over \(\bk\) may be converted to an integral:
\begin{equation*}
\sum_\bk \to 2 \frac{V}{(2\pi)^3} \int d\Omega \int_0^\infty dk \,k^2.
\end{equation*}
By introducing the matrix
\begin{equation}\label{eq:S}
S_{j'j} = \frac{3}{4 \pi} \sum_i \int d\Omega (\mathbf{p}_{ij'}\cdot \ephat )^*(\mathbf{p}_{ij}\cdot \ephat )
\end{equation}
and changing variables to integrate on \(\nu_\bk = ck/n\), the equation of motion (Eq.~\ref{eq:eqmotion}) becomes
\begin{equation*}
\begin{aligned}
\dot{b}_{j'}(t) =  - \frac{1}{6 \pi^2}  & \frac{ n^3}{ \hbar \epsilon c^3 } \cdot  \sum_{j} S_{j'j} \int_o^\infty d\nu_\bk \nu_\bk^3 \int_0^t dt'\\
&  e^{-i (\omega - \nu_\bk)(t-t')} e^{i (\Delta_{ij}t'-  \Delta_{ij'}t )} b_j(t').
\end{aligned}
\end{equation*}

Since the integral over \(t\) is only appreciable when \(\omega \sim \nu_\bk\), \(\nu_\bk^3\) may be replaced with \(\omega^3\) in the integrand and the lower frequency limit may be replaced by \(-\infty\)~\cite{Scully,MesoQO}. Using the delta function identity,
\begin{equation*}
\int_{-\infty}^\infty d\nu_\bk e^{-i (\omega - \nu_\bk)(t-t')} = 2 \pi \delta(t-t').
\end{equation*}
and
\[ \int_0^t dt' \delta(t-t') f(t') = \frac{1}{2} f(t),\]
we arrive at the differential equations in the desired form:

\begin{equation}\label{eq:result}
\dot{b}_{j'}(t) = -\frac{\alpha}{2} \sum_j S_{j'j} b_j(t) e^{i \Delta_{jj'}t } \, , \quad \alpha = \frac{1}{4 \pi \epsilon}  \frac{4 \omega^3 n^3}{3 \hbar c^3}  .
\end{equation}

The matrix \(S_{j'j}\) can be computed from \(\mathbf p_{ij}\) in a simple way. Using the parameterization
\[
\ephat = \sin \theta \cos \phi \, \bm{\unit x} +  \sin \theta \sin \phi \, \bm{\unit y}+  \cos \theta  \, \bm{\unit z}.
\]
for the polarization, the integrand \eqref{eq:S} contains a sum of integrals of the form
\[ I_{nm} =  \int d\Omega (\bm{\unit{x}_n} \cdot \ephat  ) ( \bm{\unit{x}_m} \cdot \ephat ) \]
where \(\bm{\unit{x}_n}\) is a unit vector. Performing the angular integrals, this becomes
\[ I_{nm} = \frac{4 \pi}{3} \delta_{nm} .\]
Thus we arrive at a convenient shorthand for computing \(S_{j'j}\) given the dipole matrix:
\begin{equation*}
S_{j'j} = \sum_i \mathbf{p}_{ij'}^* {\cdot}\mathbf{p}_{ij}
\end{equation*}
where the dot product is evaluated using \( \unit{x}_n \cdot \unit{x}_m = \delta_{nm} \). This shows the angular integral can be replaced with a simple dot product.

In matrix language, the differential equation governing excited state probability amplitudes for degenerate excited states is
\begin{equation*}
\frac{d\mathbf b}{dt} = - \frac{\alpha}{2} S \mathbf b, \quad S =  \mathbf{p}^\dagger{\cdot}\mathbf{p}.
\end{equation*}
By choosing a basis for the excited states in which \(S\) is diagonal, the decay of each state is uncoupled from the others. Thus we see that the the eigenstates of \(\alpha S\) decay independently at spontaneous emission rates equal to the eigenvalues of \(\alpha S\).

\begin{figure*}
\includegraphics{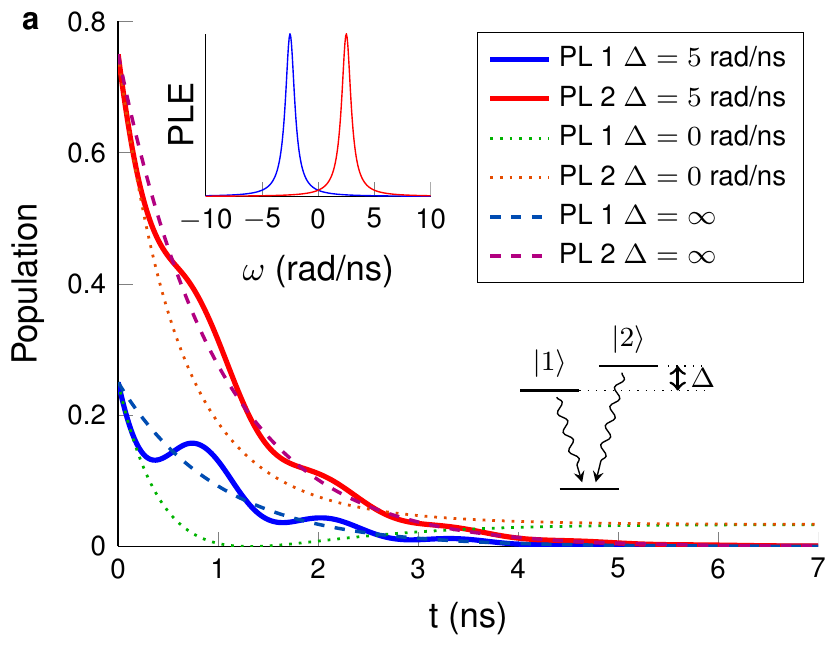}\includegraphics{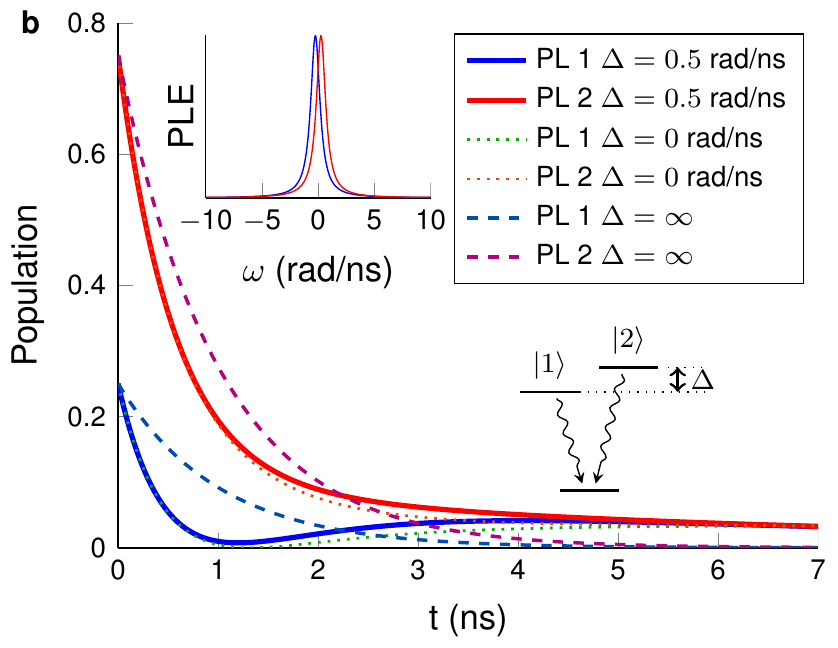}
\caption{Time dependence of excited state population depending on detuning between two excited states that can interfere. \(\tau=1\)~ns. For a detuning \(\Delta\) much larger than the spontaneous emission rate \(1/\tau\), the population of the two excited states decay independently. When \(\Delta=0\), the lifetimes are modified by interference of the two recombination pathways, and the system progresses towards a dark state. The lifetimes of states in the absence of interference are 1~ns for both transitions. The initial wave function is \(\ket \psi = \frac{1}{2} \ket 1 + \frac{\sqrt{3}}{2} \ket 2 \). \textbf{a.} At intermediate \(\Delta=5\) rad/ns, some interference is seen in the emission, oscillating around the behavior expected for two independent subsystems. \textbf{b.} When the transitions are nearly resonant, the emission follows the behavior for degenerate excited states.} 
\label{fig:detunedUpperStates}
\end{figure*}

This method can be used to compare spontaneous emission rates between different manifolds of excited states. If the states are split in energy by a large amount compared to the radiative lifetime, the fast oscillating term in Eq.~\ref{eq:result} will result in non-degenerate excited states becoming uncoupled. As an example, we solved Eq.~\ref{eq:result} for two excited states and one ground state, whose lifetimes can be modified if the excited levels are degenerate. The differential equations governing the excited state probability amplitudes are
\begin{equation*}
\frac{d}{dt}
\left(
\begin{array}{c}
  {b}_1   \\
   {b}_2   
\end{array}
\right) = -\frac{1}{2 \tau}
\left(
\begin{array}{cc}
  1 & e^{i \Delta t}   \\
  e^{-i \Delta t} & 1 \\
\end{array}
\right)
\left(
\begin{array}{c}
  {b}_1   \\
   {b}_2   
\end{array}
\right)
\end{equation*}
The solution is plotted in Fig.~\ref{fig:detunedUpperStates} for \(\tau=1\)~ns and various detunings \(\Delta\). 
If the transitions are well resolved, the time dependence of the system follows that of two independent subsystems (Fig.~\ref{fig:detunedUpperStates}a). On the other hand, if there is significant overlap between the Lorentzian line shapes, interference effects modify the radiative lifetime. In this toy model, near-degeneracy results in the existence of a dark state and long lived excited state population~(Fig.~\ref{fig:detunedUpperStates}b).

Therefore, to find the spontaneous emission rates for different non-degenerate sets of states, it is only necessary to calculate the eigenvalues of \(S\) within each degenerate subspace. When comparing eigenrates between non-degenerate manifolds, the existence of \(\omega^3\) in the pre-factor \(\alpha\) modifies the lifetimes. For excited state splittings of 10s of GHz (as for \azerox) at optical frequencies, this leads to a correction of a part in \(10^4\), which can often be neglected.

\section{Density matrix model}\label{sec:model}

The time evolution of the excited state density matrix \(\rho\) is described by
\begin{equation}\label{eq:equationOfMotion}
 \frac{d\rho}{dt}= \frac{1}{i \hbar} [H_o,\rho] - \frac{1}{2} \left\{\rho, \alpha \mathbf p^\dagger{\cdot}\mathbf p \right\} + L(\rho)
 \end{equation}
 where \(H_o\) is a diagonal matrix of the energies of the excited state energies, \(\mathbf p\) is the dipole matrix with elements \(\mathbf p_{ij} = \bra{g,i} \boldsymbol \mu \ket{e,j} \) and \(L(\rho)\) is the Linbladian operator. The first term describes unitary evolution, the second term spontaneous emission and the third excited state relaxation and decoherence. In this section, we describe the construction of this model.

A population transfer out of the system is mathematically identical to the spontaneous emission process in the excited state subspace. Population reduction can be accomplished mathematically using an anti-commutator \(\{A,B\}=AB+BA\)~\cite{Scully}. The spontaneous emission process is characterized by the decay of diagonal density matrix terms in the basis of the spontaneous emission eigenstates:
\begin{equation}\label{eq:rad}
\left( \frac{d\rho}{dt} \right)_\text{radiative} =  - \frac{1}{2} \sum_j \left\{\rho, \gamma_j \ket{\phi_j} \bra{\phi_j} \right\} ,\end{equation}
where \(\gamma_j\) and \(\ket{\phi_j}\) are the spontaneous emission eigenvalues and eigenstates of \( \alpha S\). Using the fact that any operator is diagonal in the basis of it's eigenvectors, this becomes
\begin{equation*}\label{eq:rad}
\left( \frac{d\rho}{dt} \right)_\text{radiative} =  - \frac{\alpha}{2} \left\{\rho, S \right\},
\vspace{2mm}
\end{equation*}
Numerically, it is an advantage to include spontaneous emission in this way as the ground states do not need to be included in the density matrix. For \azerox, the excited state density matrix has 56 differential equations (ignoring off diagonal terms between non-degenerate excited states), whereas a treatment using the full density matrix including ground states would have 120.

Phenomenological relaxation between the excited states is included as a population transfer and decoherence. The excited state relaxation rate \(R_{ij}\) from state \(i\) to \(j\) arises from coupling between the spins and their environment. This model includes an intralevel relaxation rate between states in a degenerate manifold, and an interlevel rate between different manifolds. For \(i\) and \(j\) in different manifolds, the rates are modulated by the energy difference between the initial and final states,
\[ R_{ij} \propto R_o e^{ - \frac{(E_i-E_j)}{2k_B T}} ,\]
where \(R_o\) is the interlevel relaxation rate. This correctly reproduces the fact that in equilibrium the ratio of populations in different states is given by a Boltzmann factor.
Many of the rates \(R_{ij}\) from states in the same irreducible representation \(\Gamma_n\) can be shown to be the same by symmetry. For the purposes of this model, we assumed that all states within a given irreducible representation \(\Gamma_n\) have the same phenomenological relaxation rate.

Phenomenological relaxation affects both the on- and off-diagonal elements of the density matrix. The total rate of population leaving (\(L_i\)) or entering (\(E_i\)) state \(i\) is
\[ L_i = \rho_{ii} \sum_j R_{ij}, \quad E_i = \sum_j \rho_{jj} R_{ji} .\]
This population relaxation also causes a decay of the off-diagonal terms in the density matrix. This can be accomplished with the anti-commutator:
\begin{equation}\label{eq:rad}
L(\rho) =  - \frac{1}{2} \sum_j \left(  L_i \delta_{ij} \rho_{jk} + \rho_{ij} L_j \delta_{jk} \right) + E_i \delta_{ik}
\end{equation}
where the second term enforces conservation of excited state population.

\section{Fit of time resolved data to model}\label{sec:uncert}

We numerically integrated the equation of motion Eq.~\ref{eq:equationOfMotion} to find the excited state density matrix as a function of time \(\rho(t)\). In the case that \(\Gamma_{n}\) is excited with polarization \(\hat{\epsilon}_\text{i}\), the initial density matrix in the \(\Gamma_n\) subspace is
\begin{equation}\label{eq:rhoExcite}
\rho_e^{(\Gamma_n)}(t=0) = M_{\epsilon_\text{i}}^{(\Gamma_n)\dagger}  M_{\epsilon_\text{i}}^{(\Gamma_n)},
\end{equation}
where \(M_{\epsilon_\text{i}}^{(\Gamma_n)} = \mathbf{p}^{(\Gamma_n)} \cdot \ephat_\text{i}\) and all other terms of the density matrix are zero.
From the solution of the density matrix as a function of time \(\rho_e(t)\), the PL emitted from \(\Gamma_n\) at polarization \(\hat{\epsilon}_\text{f}\) as a function of time is
\begin{equation}\label{eq:PLintsupp}
\text{PL}^{(\Gamma_n)}(t) = \text{tr} \left( M_{\epsilon_\text{f}}^{(\Gamma_n)} \rho_e^{(\Gamma_n)}(t) M_{n\epsilon_\text{f}}^{(\Gamma_n)\dagger}             \right).
\end{equation}
These PL curves predict the time dependence of \azerox emission under different excitation conditions.

\begin{figure*}
\includegraphics{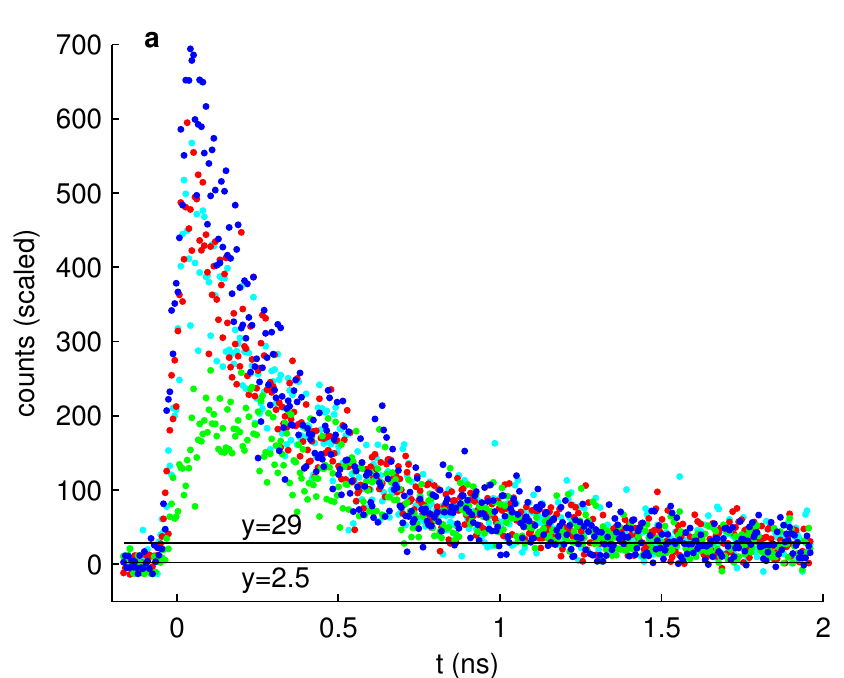}\includegraphics{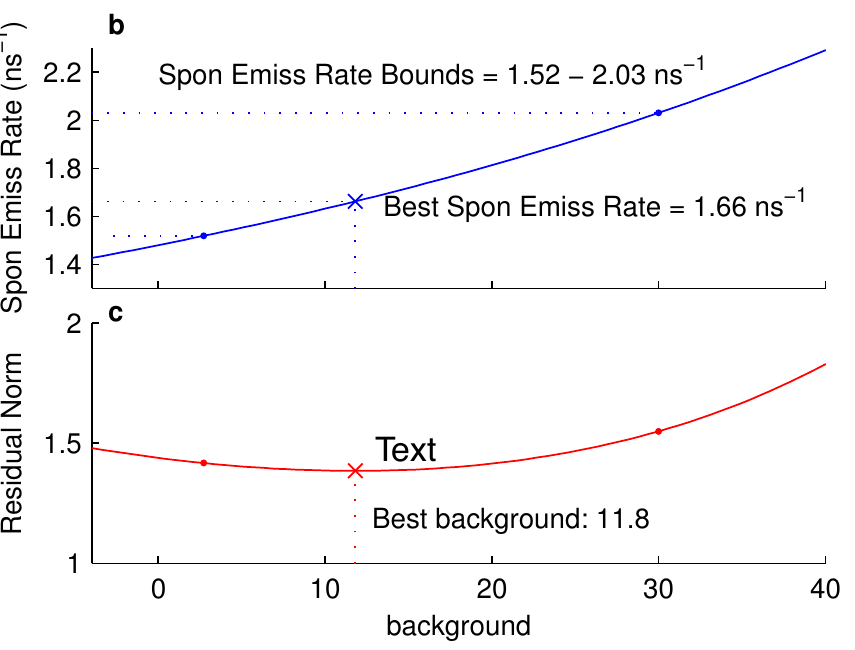}
\caption{\textbf{a.} Raw data before binning used for time-resolved experiment. Before the pulse arrives, the mean background level is \( 2.5 \pm  0.6\) counts. At the end of the trace at 2 ns, the mean level is \(29 \pm 6 \), which is somewhat higher than the theory would predict. \textbf{b-c:} Best fit spontaneous emission rate as a function of background level in the model. The best fit background value lies in between the limits chosen by inspection in a. These limits are used to find the uncertainty in the spontaneous emission rate due to the background level.
}
\label{fig:rawData}
\end{figure*}

The model was fit to the data using a weighted least-squares residual due to the Poisson distributed nature of photon counting data~\cite{Turton2003}. Temperature was included as a fit parameter because in individual fits with \(T=1.9\) K, the best fit \G{3} \(\to\) \G{5} relaxation rate depended on the excitation state. This implies that the effective exciton temperature was higher than 1.9 K, consistent with experiments on free excitons in GaAs where the effective temperature of free excitons was found to be somewhat higher than that of the bath~\cite{Ruhle1978}. 

We tested modifying the relaxation rate matrix \(R_{ij}\) so that electron spin flips can occur. In this case, the model also fits the data with different rates of interstate relaxation: interlevel relaxation (0.45~$\text{ns}^{-1}$), intralevel \G{3} relaxation (1.2~$\text{ns}^{-1}$) and intralevel \G{5} relaxation (0.71~$\text{ns}^{-1}$).
Because we obtain good fits with either electron spin flips allowed or disallowed, the experiment is not sensitive to the rate of electron spin flips. However, the best fit spontaneous emission rate constant was the same 1.48~$\text{ns}^{-1}$ regardless of whether electron spin flips are allowed.

In order to estimate the uncertainty in the the measured spontaneous emission rate, we characterized the uncertainty due to  random Poisson noise in the measurement data as well as systematic error due to the uncertainty in model parameters (e.g. pulse arrival time, background level).
The largest uncertainty in measured spontaneous emission rate arises from the uncertainty of background level (Tab.~\ref{tab:uncert}). The raw data shown in Fig.~\ref{fig:rawData} shows that there is a long lived emission above the background. This long lived emission may indicate some long lived state, e.g. exciton hopping into a metastable state and subsequent slow repopulation. 
Due to the uncertainty of the true background value, the \azerox best fit parameters acquire some uncertainty. Choosing a higher background level results in a faster best fit spontaneous emission rate, as the effective curvature of the decay becomes greater (Fig.~\ref{fig:rawData}). 
We take the confidence interval of the background level to be 0 to 30, 
this produces an uncertainty in spontaneous emission rate of \(\pm 0.26 \text{ ns}^{-1}\)  (Fig.~\ref{fig:rawData}).
Another way to estimate this uncertainty would be to incorporate a metastable excitonic level in the model. Because this introduces the danger of overfitting the model with too many adjustable parameters, we used the background level as a proxy for the uncertainty introduced by possible metastable states. 

\begin{table}[htdp]
\caption{Summary of uncertainties in measurement of the spontaneous emission rate.}
\begin{center}
\begin{tabular}{c c}
Effect & Uncert. (\(\pm\)) in Spon. Emission\\
\hline
Background Level & 0.26 $\text{ ns}^{-1}$ \\
Data Cut Off & 0.15 $\text{ ns}^{-1}$ \\
Poisson Noise & 0.0082 $\text{ ns}^{-1}$ \\
5\% Laser Power Fluctuations & 0.0047 $\text{ ns}^{-1}$ \\
\end{tabular}
\end{center}
\label{tab:uncert}
\end{table}%

Next, we investigated whether changing the maximum number of data points collected (1-2 ns) modified the best fit spontaneous emission rate. In fitting the data, there is a somewhat arbitrary choice of when additional data points at longer times no longer improve the fit. Within reasonable choices of the data time cut-off of 1-2~ns, we found that the best fit spontaneous emission rate changed from 1.36 to 1.66 $\text{ns}^{-1}$. This level of uncertainty is lower than that present from the unknown background level.

%
%
Next, we used a Monte Carlo simulation to determine the uncertainty in spontaneous emission due to Poisson noise. The raw data was used as the mean for new Poisson-distributed datasets. The model was fit to the new random datasets using the same weighted least-squares algorithm. The standard deviation of the resultant spontaneous emission rates is 0.0082 $\text{ns}^{-1}$. 

Monte Carlo simulations were also employed to calculate the uncertainty due to laser power fluctuations between experimental runs. The photon counting data was modulated by random 5\% power fluctuations and passed through the least squares algorithm. We found the standard deviation of best fit spontaneous emission rates to be 0.0047 $\text{ns}^{-1}$ due to power fluctuations of the laser. These simulations demonstrate that the measurement is robust against Poisson noise and laser power fluctuations.

In summary, we have found that the spontaneous emission rate for \azerox lies within the range 1.36 to 2.03 $\text{ns}^{-1}$. This corresponds to a lifetime constant in the range of 0.49 to 0.74 ns.


\begin{table*}[htbp]
\includegraphics{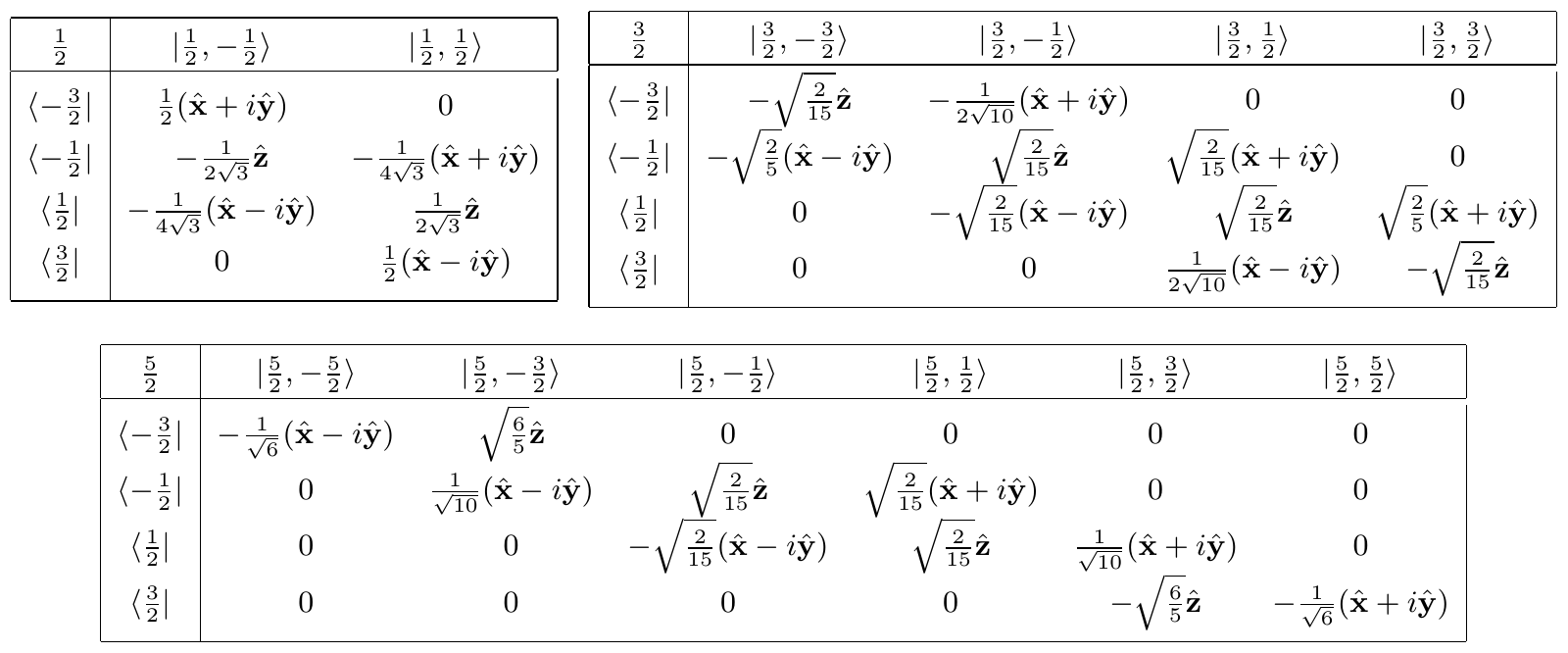}
\caption{Normalized dipole matrix \( \left( \mathbf{p}_{ij} \right)^\text{JJCS} = \bra{ \text{A}^0, i} \boldsymbol \mu \ket{ \text{A}^0 \text{X},j }\) for the \azerox-\azero system in the (incorrect) JJCS. \(\unit{x}\), \(\unit{y}\) and \(\unit{z}\) are unit vectors oriented along the crystallographic axes.}\label{tab:dipoleMatJJCS}
\end{table*}

\section{Polarization of PL to determine dominant coupling in \azerox}\label{sec:pol}

While the splitting of the \azerox states into three sets of states is now understood, it was at one point a subject of debate. Two theories, the \(j-j\) coupling scheme (JJCS) and the crystal-field scheme (CFS) can be used to explain some of the optical properties of \azerox~\cite{Karasyuk1998,Mathieu1984}. In both schemes, hole-hole coupling first rearranges the two \(j=\frac{3}{2}\) hole states into \(j=0\) and \(j=2\) manifolds. In the JJCS, electron-hole coupling further splits the \azerox states, resulting in \(j=\frac{1}{2}\) (arising from \(j=0\)) and \(\frac{3}{2}\),~\(\frac{5}{2}\) (from \(j=2\)). On the other hand in the CFS, GaAs crystal fields split the \azerox states into \(\Gamma_1\) (\(j=0\)) and \(\Gamma_{3},\Gamma_{5}\) (\(j=2\)).

In previous studies, the stress dependence of \azerox~\(\to\)~\azero emission was used to determine that only the CFS adequately describes the \azerox~\cite{Karasyuk1998,Mathieu1984}. Low temperature stress dependencies are challenging experiments, requiring the use of special equipment. In this section, we demonstrate that simple polarization measurements can also distinguish between the JJCS and CFS.

\begin{figure}
\includegraphics{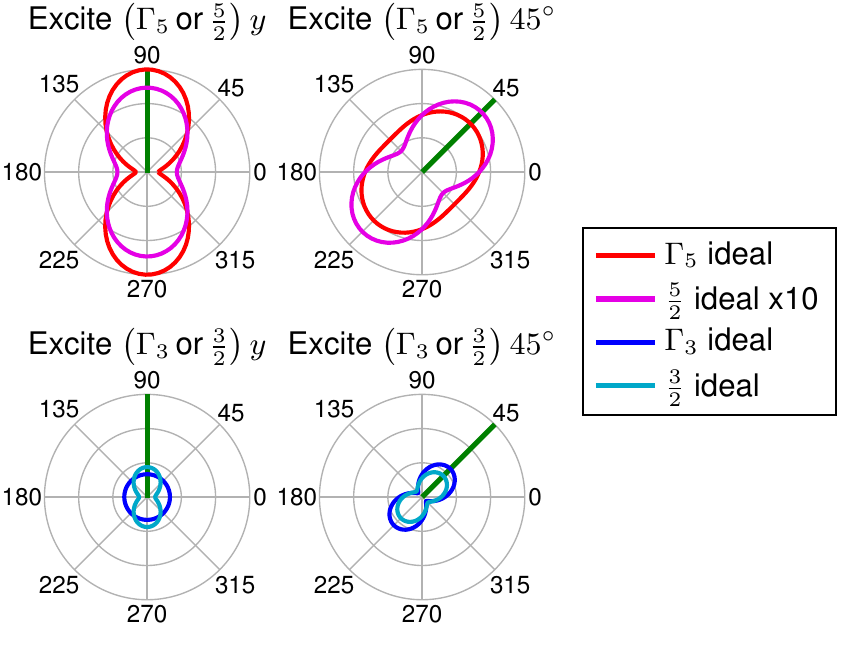}
\caption{Comparison of polarization dependence of photoluminescence for the JJCS and CFS. Qualitatively different behavior is observed when exciting \(\Gamma_3\,y\) or \( \frac{3}{2} \,y\). This difference can be used to determine the validity of the CFS for describing \azerox.}
\label{fig:compareCouplings}
\end{figure}

In order to predict the polarization dependence of \azerox in the JJCS, we first calculate the dipole matrix elements using the JJCS basis states by the procedure in Sec.~\ref{sec:dipoleMatrixElements}. The dipole matrix elements are given in Tab.~\ref{tab:dipoleMatJJCS}.

We will now calculate the polarization dependence of the PL intensity in the case of no excited state relaxation for \(\Gamma_3\) (CFS). When \azerox absorbs a photon resonant with the \(\Gamma_3\) transition, the excited state density matrix is proportional to
\begin{equation}\label{eq:rhoExcite}
\rho_e^{(\Gamma_3)} = M_{\epsilon_\text{i}}^{(\Gamma_3)\dagger} \rho_0 M_{\epsilon_\text{i}}^{(\Gamma_3)},
\end{equation}
where \(\rho_0 =\frac{1}{4} I\) is the ground state density matrix before excitation and \(M_{\Gamma_3,\epsilon_\text{i}} = \mathbf{p}_{(\Gamma_n)} \cdot \ephat_i\) are the dipole matrix elements given in Tab.~\ref{tab:dipoleMat} evaluated with the polarization \(\epsilon_i = \cos{\phi_\text{i}} \unit x + \sin{\phi_\text{i}} \unit y\) in the \(x-y\) plane.
After absorption, the part of the excited state density matrix corresponding to \(\Gamma_{3}\) is
\begin{equation*}
\rho_e^{(\Gamma_3)} = \left(\begin{array}{cccc}
\frac{1}{12} & 0 &  -\frac{i \sin(2 \phi_\text{i})}{8\sqrt{3}} & 0\\
0& \frac{1}{12} & 0 &  -\frac{i \sin(2 \phi_\text{i})}{8\sqrt{3}}  \\
\frac{i \sin(2 \phi_\text{i})}{8\sqrt{3}}   &0 & \frac{1}{12} & 0 \\
0&  -\frac{i \sin(2 \phi_\text{i})}{8\sqrt{3}}   &0 & \frac{1}{12} \\
\end{array}
\right)
\end{equation*}
with all other excited state density matrix elements equal to zero. (Here we are working in the crystal field scheme basis.) To find the amount of PL emitted with linear polarization  \(\epsilon_\text{f} = \cos{\phi_\text{f}} \unit x + \sin{\phi_\text{f}} \unit y\), we evaluate
\begin{equation*}
\text{PL}^{(\Gamma_3)} = \text{tr} \left( M_{\epsilon_\text{f}}^{(\Gamma_3)} \, \rho_e \,M^{(\Gamma_3)\dagger}_{\epsilon_\text{f}}   \right).
\end{equation*}
Simplifying, and repeating this procedure for \(\Gamma_1\) and \(\Gamma_5\), the angular dependence of polarization in the case of no excited relaxation is
\begin{align*}
\text{PL}^{(\Gamma_5)} &=  \frac{I_o}{18} [5 +  4 \cos(2 \phi_\text{i}) \cos(2 \phi_\text{f})  + \sin(2 \phi_\text{i})\sin(2\phi_\text{f}) ] \\
\text{PL}^{(\Gamma_3)} &= \frac{I_o}{36} \left[ 4 + 3 \sin ( 2 \phi_\text{i}) \sin (2 \phi_\text{f}) \right] \label{eq:G3xy} \\
\text{PL}^{(\Gamma_1)} &= \frac{I_o}{18}.
\end{align*}

On the other hand, if the JJCS dipole operator (Tab.~\ref{tab:dipoleMatJJCS}) is used, the PL from the three manifolds is
\begin{align*}
\text{PL}^{(5/2)} &=  I_o \left[ \frac{1}{36} + \frac{1}{75} \cos(2 \phi_\text{f}-2 \phi_\text{i}) \right] \\ 
\text{PL}^{(3/2)} &=  I_o \left[ \frac{533}{5760} + \frac{4}{75} \cos(2 \phi_\text{f}-2 \phi_\text{i}) \right] \\
\text{PL}^{(1/2)} &=  I_o \frac{169}{4608}.
\end{align*}
The two coupling schemes show qualitatively different angular polarization dependences, shown in Fig.~\ref{fig:compareCouplings}. By comparing with the experimental data shown in Fig.~\ref{main-fig:fits}a in the main text, we conclude that only the CFS adequately describes the angular dependence of \azerox photoemission.

\section{Photoluminescence Excitation Spectroscopy of \azerox}\label{sec:inhomo}

\begin{figure}
\includegraphics{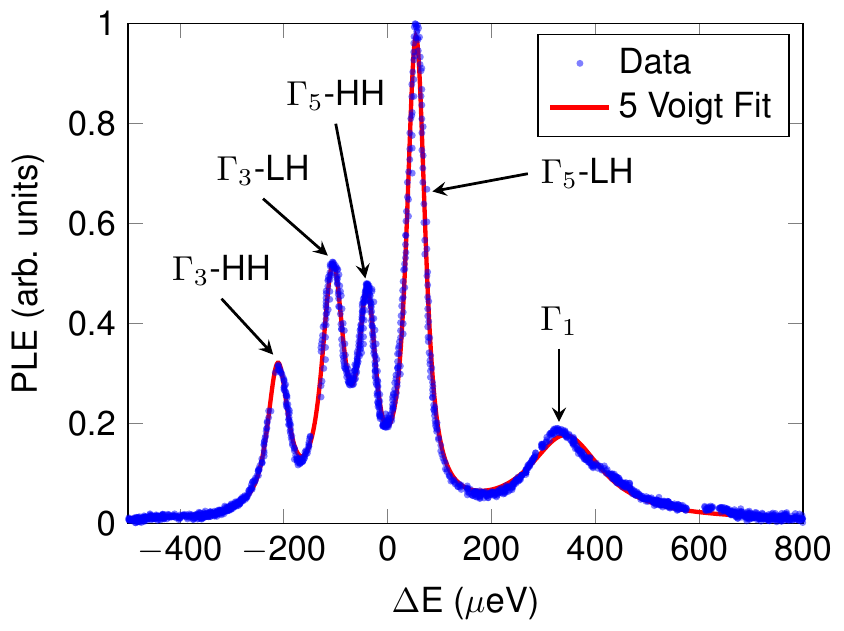}
\caption{Photoluminescence excitation spectroscopy of \azerox. The method of mounting the sample introduces strain, which splits the heavy-hole and light-hole ground states. A linear offset due to donor acceptor pair emission was subtracted for clarity. The curve was fit to a sum of five Voigt functions. 
}
\label{fig:PLE}
\end{figure}

The \azerox system is a remarkably homogeneous excitonic system. To investigate the inhomogeneous broadening of the \azerox system, we perform photoluminescence excitation (PLE) spectroscopy on a p-type GaAs sample mounted in a cold-finger cryostat at 4.2 K. The method of mounting the sample introduced some strain into the sample, which splits the heavy hole (HH) and light hole (LH) states. A narrow band ($<$10~neV) continuous-wave laser is scanned over the \azero-1s to \azerox transition while monitoring PL from \azerox to \azero-2s (Fig.~\ref{fig:PLE}).

We fit the PLE lines to a sum of five Voigt functions, the convolution of a Lorentzian and a Gaussian. The Lorentzian width is due to homogeneous effects while the Gaussian width arises from inhomogeneous broadening. In the fit, the inhomogeneous broadening is the same for all peaks. The best fit Lorentzian full width at half maximum are $(39 \pm 2)\,\mu$eV for $\Gamma_3$-HH, $(43 \pm 2)\,\mu$eV for $\Gamma_3$-LH, $(29 \pm 2)\,\mu$eV for $\Gamma_5$-HH, $(33 \pm 1)\,\mu$eV for $\Gamma_5$-LH and $(170 \pm 7)\,\mu$eV for $\Gamma_1$. The inhomogeneous broadening full width at half maximum was $(19 \pm 1)\,\mu$eV. Thus we find that \azerox is a remarkably homogeneous excitonic system.


\bibliography{toddkarin.bib}

%
%

%
%
\end{document}